\documentclass[twocolumn,preprintnumbers,amsmath,amssymb]{revtex4}
\usepackage{amsfonts}
\usepackage{epsf}
\usepackage{anysize}
\marginsize{0.8 in}{0.8 in}{1in}{1.4in}
\usepackage{float}
\usepackage{graphicx}
\usepackage{dcolumn}
\usepackage{bm}
\usepackage{amsmath}
\usepackage{amssymb}
\usepackage{epstopdf}
\usepackage{mathrsfs}
\usepackage{pstricks,pst-node,pst-text,pst-3d}
\usepackage{subfigure}
\usepackage[all]{xy}
\usepackage{color}
\usepackage{epsf}
\usepackage{anysize}
\usepackage{graphicx}
\usepackage{dcolumn}
\usepackage{bm}
\usepackage{amsmath}
\usepackage{amssymb}
\usepackage{mathrsfs}
\usepackage{pstricks,pst-node,pst-text,pst-3d}
\usepackage[all]{xy}
\usepackage{mathtools}
\usepackage{appendix}

\DeclarePairedDelimiterX\braket[2]{\langle}{\rangle}{#1 \delimsize\vert #2}

\makeatletter
\newcommand\figcaption{\def\@captype{figure}\caption}
\newcommand\tabcaption{\def\@captype{table}\caption}
\makeatother

\newcommand{\tr}{{\rm Tr}}

\newcommand{\la}{\mathcal{L}}

\newcommand{\n}{{\bf{n}}}

\begin{document}


\title{Identification of Time-varying \textit{in situ} Signals in Quantum Circuits}

\author{Xi Cao,$^{1,\dagger}$, Yu-xi Liu$^{2,3}$ and Re-Bing Wu$^{1,3,}$}
\thanks{rbwu@tsinghua.edu.cn}
\address{ \small$^{1}$Department of Automation, Tsinghua University, Beijing 100084, China\\
	\small$^{5}$Institute of Micro-Nano Electronics, Tsinghua University, Beijing 100084, China\\
	\small$^{4}$Center for Quantum Information Science and Technology, BNRist, Beijing 100084, China}


\begin{abstract}
	The identification of time-varying \textit{in situ} signals is crucial for characterizing the dynamics of quantum processes occurring in highly isolated environments. Under certain circumstances, they can be identified from time-resolved measurements via Ramsey interferometry experiments, but only with very special probe systems can the signals be explicitly read out, and a theoretical analysis is lacking on whether the measurement data are sufficient for unambiguous identification. In this paper, we formulate this problem as the invertibility of the underlying quantum input-output system, and derive the algebraic identifiability criterion and the algorithm for numerically identifying the signals. The criterion and algorithm can be applied to both closed and open quantum systems, and their effectiveness is demonstrated by numerical examples.
\end{abstract}
\maketitle

%

\section{Introduction}
\begin{figure}[]
	\centering
	\includegraphics[width=0.9\columnwidth]{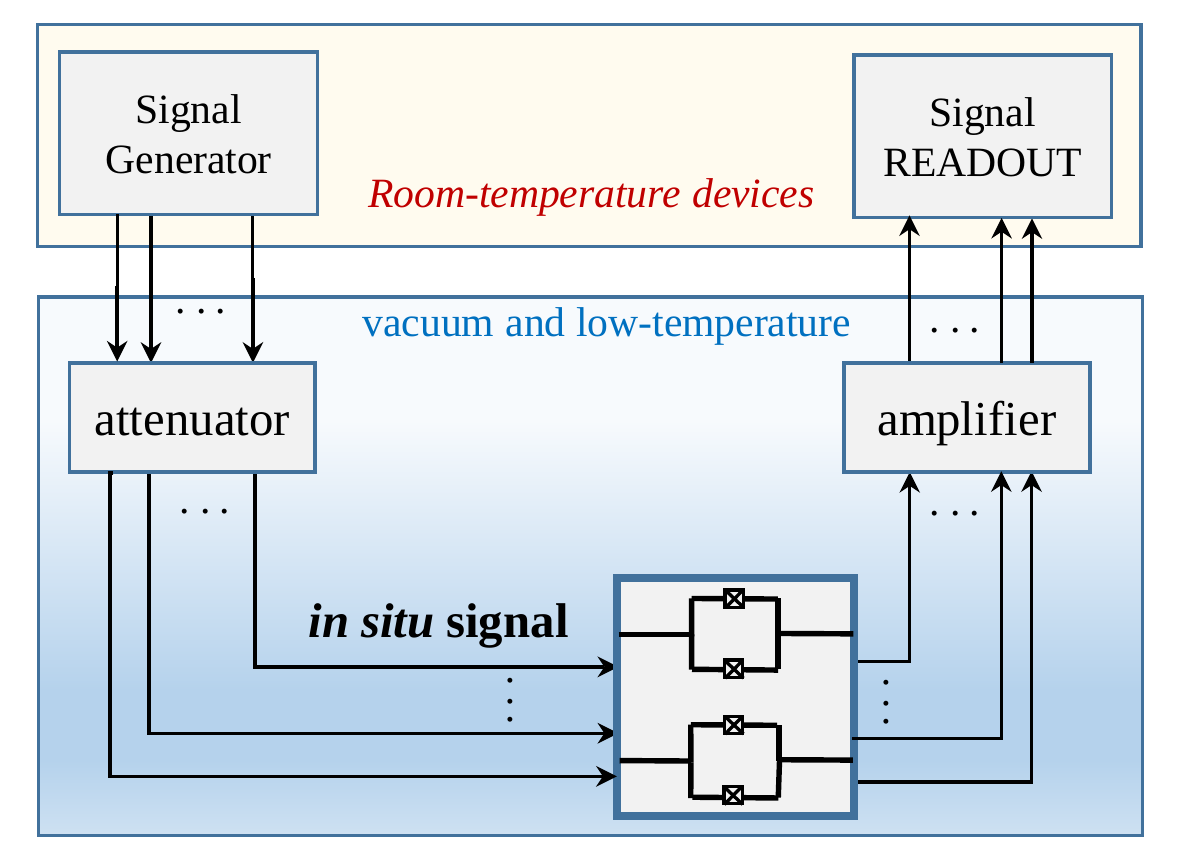}
	\caption{Schematic diagram of a superconducting quantum computing platform. The {\it in situ} signals are delivered from ambient signal generators that may experience distortion along the transmission line. The signals are fed into a quantum chip of multiple qubits, whose measurement signal are conducted out for identifying the {\it in situ} signals.}
	\label{device}
\end{figure}

The full characterization of quantum dynamics is crucial for high-precision modeling and manipulation of quantum information processing systems. In the literature, systematic studies have been casted to the identification of quantum states and operations (as known as quantum tomography) \cite{Banaszek2013,Anis_2012,Bra_2012} or quantum Hamiltonians \cite{Maruyama2012,7798641,HOU2017863,LeBris2007a,Zhang2014,Alis2004,PhysRevA.87.022324}. Most of these works focus on the estimation of constant but unknown quantities, e.g., a density or process matrix \cite{Maruyama2012,7798641,HOU2017863,LeBris2007a,Zhang2014} or some parameters in the Hamiltonian \cite{Alis2004,PhysRevA.87.022324}, based on maximum-likelihood, least-square or comprehensive sensing estimators. However, the identification of unknown time-varying signals has been rarely studied so far. Such problems broadly exist in low-temperature quantum information processing systems where {\it in situ} signals are not reachable by ambient measurement devices. For the example of superconducting quantum chips \cite{johnson2011controlling} shown in Fig.~\ref{device}, the {\it in situ} DC or AC control signals in low temperature environment always experience distortion along the attenuators and the control line \cite{PhysRevApplied.4.024012,PhysRevA.84.022307,SPINDLER201249,Glaser2015,PhysRevX.8.011030,PATTERSON2006231,article,johnson2011controlling,Cao}, but the distorted signals can not be directly acquired by the qubit readout signals.

Most {\it in situ} signals have to be indirectly extracted from a quantum probe (e.g., a qubit). For instance some {\it in situ} signals can be readout from the qubit phase variance measured by Ramsey experiments (see Section \ref{sec:ramsey}) \cite{Max,PhysRevLett.109.240504,Vion2003}, but such scheme is not generalizable to more complicated systems. More seriously, as will be shown in Sec.~\ref{sec:ramsey}, there may exist non-unique estimations among which it is uneasy to determine the true one. Therefore, whether the signals are theoretically identifiable, and how to uniquely identify them, are to be well understood.

From a system point of view, the identification of time-resolved signals from time-varying measurements can be thought of as reversing the system's input-output mapping \cite{Hirschorn1979}. Whether the signal is identifiable is equivalent to the left invertibility of the system (Chapter 5, \cite{Sussmamm1990}), i.e., the property that different inputs must produce different outputs. In parallel, the right invertibility (Chapter 5, \cite{Sussmamm1990}) is referred to as the property that any desired time-varying output can be produced by some (non-unique) input function. For examples in the classical domain, the left invertibility was applied for estimating the source of heat conduction from temperature measurements \cite{CAUDILL1994}, while the right invertibility was often used for designing tracking control of a robot along a chosen trajectory \cite{robot}. All studies collectively showed that the invertibility of a general input-output system is determined by its relative degree that can be specified by an inversion algorithm.

In the quantum domain, the left invertibility was first studied by Ong, Clark and Tarn \cite{CK1984} in a nonlinear filtering problem based on  non-demolition continuous-time measurements. This work showed that, under adequate Lie algebraic conditions, the time-varying input of a quantum system can be recovered online from the time trace of a continuously measured observable. Later on, the inversion (of right invertible systems) was also applied to the quantum control design as a reference tracking problem based on virtual feedback \cite{Gross1993,Z.M.1995,Jha2009,Brown2002,doi:10.1063/1.477857,doi:10.1063/1.1582847} or to the estimation of quantum states and Hamiltonians \cite{Alis2004,LeBris2007a}.

In this paper, we will apply the inversion-based method to the identification of time-varying \textit{in situ} signals. This can be treated as a generalization of the work of Ref.~\cite{CK1984,Clark1985}, but the measurements do not have to be non-demolitional for offline identification because one can measure the time-resolved output via ensemble average. We will also extend the invertibility criterion and inversion algorithm from the single-input-single-output case to more complicated multi-input-multi-output cases, which are useful under circumstances where multiple signals are simultaneously coupled to a multi-qubit system.

The remainder of this paper will be arranged as follows. Section \ref{sec:ramsey} shows how the ambiguity issue arises in a direct Ramsey-based identification examples, following which we propose the inversion-based method for analyzing the identifiability (i.e., invertibility) and reconstructing of the input signals. Section \ref{sec:simulation_result} provides two numerical examples, a one-qubit system with single input and a two-qubit system with multiple inputs, to show the advantage of inversion-based method, and how the singularity problem can be solved by abundant measurements. Finally, conclusions are drawn in Section V.

\section{A direct identification scheme via Ramsey interferometry}\label{sec:ramsey}
Let us start from a simple case. Suppose that the signal $u(t)$ to be identified is coupled to a single qubit probe \cite{Vion2003}, and we expect to read out $u(t)$ through the time-resolved measurement of the qubit. A simple model for the readout process can be described by the Schr\"{o}dinger equation $\dot{\psi}(t)=-iH(t)\psi(t)$, where $\psi(t)$ is the quantum state of the qubit probe and
\begin{equation}
	H(t)=u(t) \sigma_z.
	\label{im_1}
\end{equation}
Here, $\sigma_x,\sigma_y,\sigma_z$ are the standard Pauli matrices.
The qubit is prepared at the initial superposition state $\psi(0)=\frac{1}{\sqrt{2}}(|0\rangle+|1\rangle)$ and evolves as follows:
\begin{equation}\label{}
	\psi(t)=\frac{1}{\sqrt{2}}[e^{-i\theta(t)/2}|0\rangle+e^{i\theta(t)/2}|1\rangle],
\end{equation}
where the information about $u(t)$ is transferred to the accumulated phase
\begin{equation}\label{}
	\theta(t)=\int_0^t u(\tau){\rm d}\tau.
\end{equation}
In the laboratory, the phase $\theta(t)$ can be conveniently measured via a Ramsey experiment that corresponds to the expectation value of $\sigma_x$:
\begin{equation}
	y(t)=\langle\psi(t)|\sigma_x|\psi(t) \rangle=\frac{\cos\theta(t)}{2}.
	\label{im_2}
\end{equation}
Reversing the above processes, we obtain the identification formula:
\begin{equation}\label{ramsey}
	u(t)=\frac{\rm d}{{\rm d}t}\left[\pm\arccos 2y(t)+k\pi\right]= \frac{\mp\dot{y}(t)}{\sqrt{1-4y^2(t)}}.
\end{equation}

\begin{figure}[]
	\centering
	\includegraphics[width=1\columnwidth]{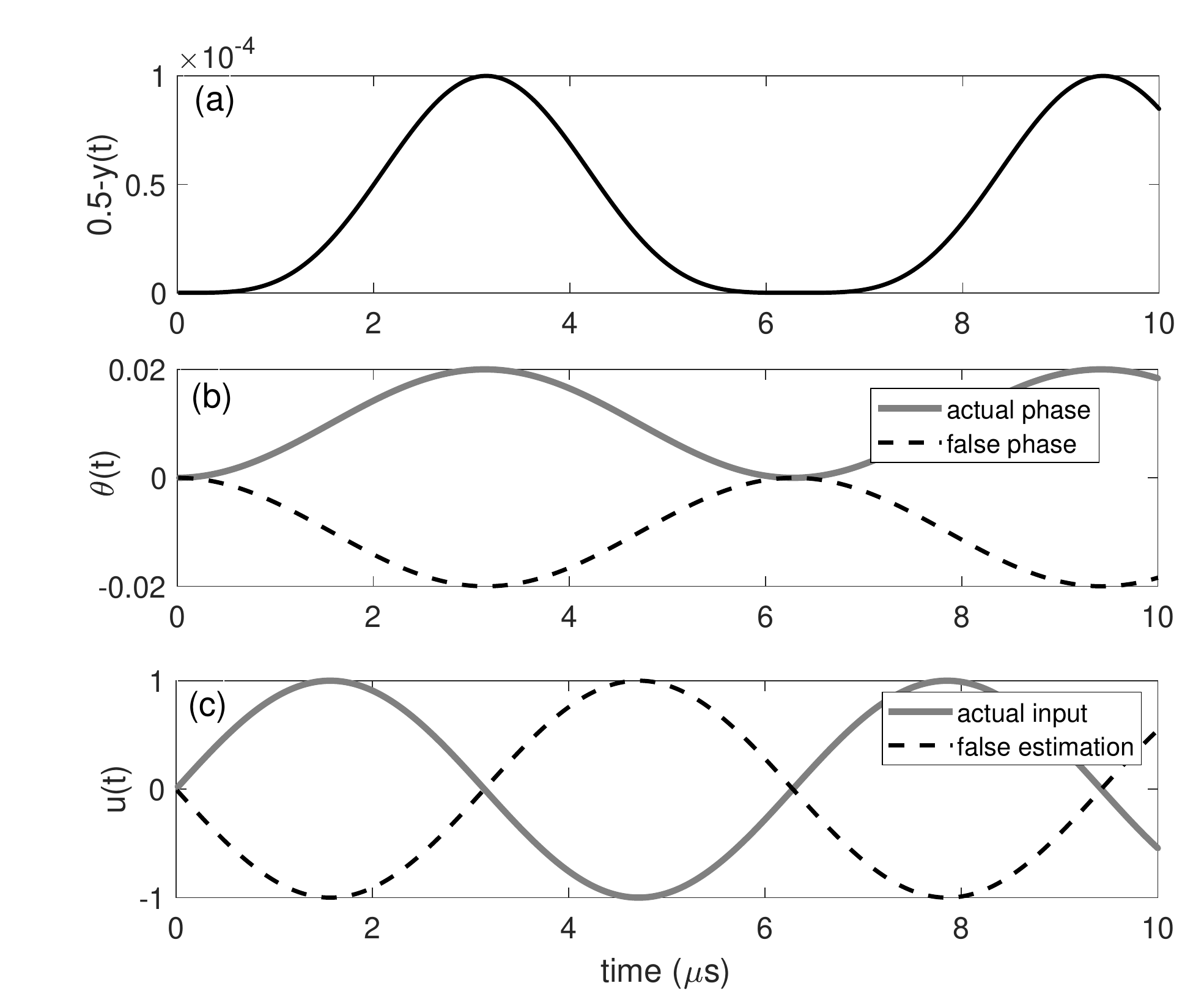}
	\caption{Ramsey experiment based identification. The measured output (a) corresponds to two different time traces of the qubit phase (b) in which one is false. The resulting identified {\it in situ} input (c) is thus undecidable. }
	\label{ramsey1}
\end{figure}

There are two issues in this identification scheme. First, the identification formula (\ref{ramsey}) provides two solutions among which one cannot determine the correct answer, because the involved cosine function is not 1-to-1. For example, as is shown in Fig.~\ref{ramsey1}, the two different signals $u(t)=\pm\sin \omega_0 t$ ($\omega_0=1$) accumulate different traces of phases that lead to the same measured output $y(t)$, and we are not able to judge whether $u(t)=\sin\omega_0t$ or $u(t)=-\sin\omega_0t$ is the real \textit{in situ} signal. Later we will see that the signal is actually identifiable but the above identification scheme is flawed.

Second, this direct identification scheme relies on the analytical solvability of the time-dependent Schr\"{o}dinger equation (\ref{im_1}), which is usually impossible when $H(t)$ and $H(t')$ do not commute for $t \neq t'$. For example, the {\it in situ} signal cannot be simply encoded into the qubit phase when there is a bias term in the qubit probe Hamiltonian:
\begin{equation}
	H(t) = \omega_0 \sigma_x + u(t) \sigma_z.
\end{equation}
This issue is even severer in multi-qubit systems that are coupled with multiple input signals. In the following section, we will show how these two problems can be resolved in a more general framework for invertibility analysis.

\section{Identification scheme by quantum system inversion}

In this section, we will take the identification problem as an inverse problem of solving the output of dynamical Schr\"{o}dinger equation, and derive the invertibility criterion as well the inversion algorithm for extracting the signals.

\subsection{Single-input single-output case}
To facilitate the following derivation, we assume that the probe system is a closed or a Markovian open system, so that the evolution can be described by:
\begin{equation}\label{eq:SISO}
	\dot{\rho}(t)= \left[\mathcal{L}_0+{u}(t)\mathcal{L}_1\right]\rho(t),
\end{equation}
where the density matrix is initially prepared at $\rho(0)=\rho_0$. The super-operators $\mathcal{L}_0$ and $\mathcal{L}_1$, which are assumed to be precisely known, are a Liouvillian ($\mathcal{L}\rho = -i[H,\rho]$ with $H$ being the interaction Hamiltonian) or a Lindbladian (i.e., $\mathcal{L}\rho = 2L\rho L^\dag-L^\dag L\rho-\rho L^\dag L$ with $L$ being an coupling operator). We expect to identify it from the time-resolved ensemble measurement
$${y}(t)=\langle O\rangle\triangleq \tr\left[\rho(t) O\right],$$
where $O$ is the corresponding observable.

The system is said to be left invertible (or functional observable) if for any admissible $u(t)\neq u'(t)$, their resulting outputs $y(t)\neq y'(t)$. To decide whether the system is invertible, we can differentiate the measurement $y(t)$:
\begin{equation}
	\dot{{y}}(t)=\langle\mathcal{L}^*_0O\rangle+\langle \mathcal{L}^*_1O\rangle{u}(t),
	\label{order}
\end{equation}
where the $\mathcal{L}^*_kO$ ($k=0,1$) represents the adjoint operation of $\mathcal{L}_{k}$ on the observable $O$, i.e., $\tr\left[(\mathcal{L}\rho)O\right]=\tr\left[\rho(\mathcal{L}^*O)\right]=\langle \mathcal{L}^* O\rangle$.

If it happens that $\langle\mathcal{L}^*_1O\rangle=0$, the differentiation can be repeated for $\alpha$ times until $\langle\mathcal{L}_1^*(\mathcal{L}_0^*)^{\alpha-1}O\rangle \neq 0$, which gives
\begin{equation}\label{order_alpha}
	y^{(\alpha)}(t)= \langle(\mathcal{L}_0^*)^{\alpha}O\rangle+\langle \mathcal{L}_1^*(\mathcal{L}_0^*)^{\alpha-1}O\rangle{u}(t).
\end{equation}
Then, we can formally write
\begin{equation}
	{u}(t)={\phi}[\rho(t),y^{(\alpha)}(t)]=\frac{y^{(\alpha)}(t)-\langle (\mathcal{L}_0^*)^{\alpha}O\rangle}{\langle\mathcal{L}_1^*(\mathcal{L}_0^*)^{\alpha-1}O\rangle}, \label{eq:InverseReadout_simple}
\end{equation}
and replace it back to Eq.~(\ref{eq:SISO}), which leads to the differential equation
\begin{equation}
	\dot{\rho}(t)=\left\{\mathcal{L}_0+{\phi}[\rho(t),y^{(\alpha)}(t)]\mathcal{L}_1\right\}\rho(t).
	\label{quantum3}
\end{equation}
This nonlinear equation forms the inverse system of (\ref{eq:SISO}) because $y(t)$ becomes the input function while the original input ${u}(t)$ becomes the output through Eq.~(\ref{eq:InverseReadout_simple}).

The index $\alpha$, which is called the relative degree of the quantum system, indicates that $u(t)$ affects $y(t)$ via its $\alpha$th-order time-derivative. It can be proven that the system is invertible if and only if and only $\alpha$ is a finite integer \cite{CK1984}.

The above process also provides an inversion algorithm that extracts $u(t)$ by numerically solving the inverse system (\ref{quantum3}) from the known measurement data ${y}(t)$ and the prepared initial state $\rho(0)=\rho_0$. Note that the condition $\langle\mathcal{L}_1^*(\mathcal{L}_0^*)^{\alpha-1}O\rangle \neq 0$ is hardly verifiable because $\rho(t)$ is not analytically solvable. In practice, we can relax this condition to the operator $\mathcal{L}_1^*(\mathcal{L}_0^*)^{\alpha-1}O$ instead of its expectation value, i.e., the system's relative degree is $\alpha$ if $\mathcal{L}^*_0(\mathcal{L}^*_1)^k{O}$ vanishes for $k=0,\dots,\alpha-1$ but not for $k=\alpha$. Under this condition, the input signals are identifiable at least on a non-empty time interval as long as $\langle\mathcal{L}_1^*(\mathcal{L}_0^*)^{\alpha-1}O\rangle$ is nonzero at $t=0$. It is possible that $\langle\mathcal{L}_1^*(\mathcal{L}_0^*)^{\alpha-1}O\rangle$ crosses zero at some nonzero time, which makes Eq.~(\ref{eq:InverseReadout_simple}) singular, the multiple solutions may exist after this time instant \cite{Jha2009}. Therefore, the relaxed algebraic condition is only necessary for invertibility.

Let us revisit the example discussed in Section \ref{sec:ramsey}. The derivation of $y(t)=\langle \sigma_x\rangle$ yields
\begin{equation}\label{}
  u(t)=-\frac{\dot{y}(t)}{\langle\psi(t)|\sigma_y|\psi(t)\rangle},
\end{equation}
which indicates that the system's relative degree is 1 and hence the system is invertible at least on a non-empty time interval as long as $\langle\psi(t)|\sigma_y|\psi(t)\rangle\neq 0$ at $t=0$. Therefore, the failure of Ramsey-based scheme is due to the improperly designed algorithm, but not due to the loss of the system's invertibility.

In comparison, we can think of the case that the measurement is chosen to be $y(t)=\langle\psi(t)|\sigma_z|\psi(t)\rangle$. It can be verified that $y^{(\alpha)}(t)\equiv0$ for any integer $\alpha$, i.e., the system is not invertible because the relative degree is infinite. In such case, there exists no algorithms by which $u(t)$ can be uniquely identified from $y(t)$.

However, when there is a bias term in the qubit probe Hamiltonian, as follows:
\begin{equation}
\dot{\psi}(t)=-i\left[\omega_a \sigma_x+u(t) \sigma_z\right]\psi(t),
\label{simple}
\end{equation}
we can differentiate $y(t)$ twice to obtain:
\begin{equation}
u(t)=\frac{\omega_a\langle \psi(t)|\sigma_z|\psi(t) \rangle-\omega_0^{-1}\ddot{y}(t)}{\langle\psi(t)| \sigma_x|\psi(t) \rangle}.
\end{equation}
Thus, this system becomes invertible with relative degree $\alpha=2$ under the same measurement.
\subsection{Multi-input Multi-output case}
Suppose that the quantum system has multiple input signals that are to be identified from multiple time-resolved measurements, as follows:
\begin{equation}\label{eq:MIMO}
\dot{\rho}(t)= \left[\mathcal{L}_0+\vec{u}(t)\cdot\vec{\mathcal{L}}_c\right]\rho(t),
\end{equation}
in which \textit{in situ} signals $\vec{u}(t)=[u_1(t),\cdots,u_m(t)]^T$ are coupled to the system via Liouvillians (or Lindbladians) $\vec{\mathcal{L}}_c=(\mathcal{L}_1,\cdots,\mathcal{L}_m)$. The dot product is referred to as the inner product $\vec{u}(t)\cdot\vec{\mathcal{L}}_c=\sum_{k=1}^m u_k(t)\mathcal{L}_k$.
We expect to identify these signals from $n$ measurements $\vec{y}(t)=[y_1(t),\cdots,y_n(t)]^T$ with $y_\ell(t)=\langle O_\ell\rangle$ being the expectation value of the corresponding observable $O_\ell$.

Similarly, the quantum system is said to be invertible if for any two different input vectors $\vec{u}(t)\neq \vec{u}'(t)$, the resulting output vectors $\vec{y}(t)\neq \vec{y}'(t)$. According to Eq.~(\ref{eq:MIMO}), we differentiate $\vec{y}(t)$ and obtain:
\begin{equation}
\dot{\vec{y}}=\langle \mathcal{L}^*_0\vec{O}\rangle+\langle \vec{\mathcal{L}}^*_c\vec{O}\rangle\vec{u},
\label{order11}
\end{equation}
where
\begin{equation}
\langle{\vec{\la}_c^*}\vec{O}\rangle=\left(
                                       \begin{array}{c}
                                         \langle\vec\la_c^*{O}_1\rangle \\
                                         \vdots \\
                                         \langle\vec\la_c^*{O}_n\rangle \\
                                       \end{array}
                                     \right),
\end{equation}
with
\begin{equation}\label{}
  \langle\vec\la_c^*{O}_k\rangle=\left(\langle\la_1^*O_k\rangle,\cdots,\langle\la_m^*O_k\rangle \right)
\end{equation}
for $k=0,1,2,\cdots,m$. Similarly, it is hard to evaluate the expectations, we analyze the corresponding operators.
We say that operator arrays $\vec\la_{c}^* O_{i_1},\cdots,\vec \la_{c}^* O_{i_p}$ are linearly independent if
$$\lambda_1\vec\la_{c}^* O_{i_1}+\cdots+ \lambda_{p}\vec\la_{c}^* O_{i_p}\neq 0$$
for any nonzero real numbers $\lambda_1, \cdots, \lambda_{p}$, and the rank of a group of operator arrays is referred to as the maximal number of mutually independent arrays in them. If there exist $m$ mutually linear independent arrays among $\vec{\mathcal{L}}_c^*O_1,\cdots,\vec{\mathcal{L}}_c^*O_n$, i.e., ${rank} (\vec{\la}^*_c\vec{O})=m$ then the signals can be formally calculated as a least-square solution of (\ref{order11}):
\begin{align}
\vec{u}(t)=&\vec{\phi}[\rho,\dot{\vec{y}}]\nonumber \\
=&\left[\langle{\vec{\la}_c^*}\vec{O}\rangle^T\langle{\vec{\la}^*_c}\vec{O}\rangle\right]^{-1}\langle{\vec{\la}^*_c}\vec{O}\rangle\left[\dot{\vec{y}}-\langle\la^*_0\vec{O}\rangle\right], \label{eq:InverseReadout}
\end{align}
This formula is then replaced back to Eq.~(\ref{eq:MIMO}) to obtain the following inverse system
 \begin{equation}
 \dot{\rho}(t)=\left\{\la_0+\vec{\phi}[\rho(t),\dot{\vec{y}}(t)]\cdot \vec{\la}_c\right\}\rho(t).
 \label{quantum4}
 \end{equation}

Similar to the single-input-single-output systems, the expectation $\langle{\vec{\la}_c^*}\vec{O}\rangle$ may become rank deficient at some time instance even when the operator rank condition ${rank}\left(\vec{\la}^*_c\vec{O}\right)=m$ is satisfied. The affection of singularity on the identification process will be discussed in the following simulation section. Moreover, the operator rank of ${\vec{\la}_c^*}\vec{O}$ may also be lower than $m$, under which circumstance Eq.~(\ref{eq:InverseReadout}) has no unique solutions for all time $t$. In such case, we need to extract $\vec{u}(t)$ via higher-order derivatives of $\vec{y}(t)$. To do this, we first divide $\vec{O}=(\vec{O}_1,\tilde{O}_1)$ such that
$$rank\left[\vec{\mathcal{L}}_c^*\vec{O}_1\right]=rank\left[\vec{\mathcal{L}}_c^*\vec{O}\right],$$
and $\vec{\mathcal{L}}_c^*\tilde{O}_1$ is linearly dependent with the arrays of $\vec{\mathcal{L}}_c^*\vec{O}_1$, i.e., there exists a matrix $V_{11}$ such that $\vec{\mathcal{L}}_c^*\tilde{O}_1=V_{11}\vec{\mathcal{L}}_c^*\vec{O}_1$.

Let $\vec{y}_1=\langle \vec{O}_1\rangle$ and $\tilde{y}_1=\langle \tilde{O}_1\rangle$. We differentiate them
\begin{eqnarray}
\langle \mathcal{L}_0^*\vec{O}_1\rangle+\langle \vec{\mathcal{L}}_c^*\vec{O}_1\rangle\vec{u}&=&\dot{\vec{y}}_1 \\
\langle \mathcal{L}_0^*\tilde{O}_1\rangle+\langle \vec{\mathcal{L}}_c^*\tilde{O}_1\rangle\vec{u}&=&\dot{\tilde{y}}_1,
\end{eqnarray}
and then eliminate $\vec{u}$ in the second equation using the relation $\vec{\mathcal{L}}_c^*\tilde{O}_1=V_{11}\vec{\mathcal{L}}_c^*\vec{O}_1$, which gives $\bar{y}_2=\langle \bar{O}_2\rangle$, where
$$\bar{O}_2={\la}_0^*\tilde{O}_1-V_{11}{\la}_0^*\vec{O}_1,\quad\bar{y}_2=\dot{\tilde{y}}_1-V_{11}\dot{\vec{y}}_1.$$
This equation can further differentiated to produce a new group of linear equations of $\vec{u}$:
\begin{eqnarray}
\langle \mathcal{L}_0^*\bar{O}_2\rangle+\langle \vec{\mathcal{L}}_c^*\bar{O}_2\rangle\vec{u}&=& \dot{\bar{y}}_2 .
\end{eqnarray}
If $\vec{\mathcal{L}}_c^*\vec{O}_1$ and $\vec{\mathcal{L}}_c^*\bar{O}_2$ includes $m$ linearly independent rows of operators, we can let $\vec{O}_2=\bar{O}_2$ and halt the process. Otherwise, we can do the same operation on $\bar{O}_2$ by separating its linearly independent part. Inductively, if the system is invertible, we can obtain a transformation $\vec O'=V\vec O=(\vec{O}_1,\cdots,\vec{O}_\alpha)^T$ of the observables after repeatedly doing the above differentiation process, which leads to the following group of equations:
\begin{eqnarray*}
\langle \mathcal{L}_0^*\vec{O}_1\rangle+\langle \vec{\mathcal{L}}_c^*\vec{O}_1\rangle\vec{u} &=&f_1[\vec{y}^{(1)},\cdots,\vec{y}^{(\alpha_r)}] \\
&\vdots&\\
\langle \mathcal{L}_0^*\vec{O}_\alpha\rangle+\langle \vec{\mathcal{L}}_c^*\vec{O}_\alpha\rangle\vec{u}&=&f_n[\vec{y}^{(1)},\cdots,\vec{y}^{(\alpha_r)}] ,
\end{eqnarray*}
in which
\begin{equation}\label{premise}
  rank\left(
           \vec{\mathcal{L}}_c^*\vec{O}'
      \right)=m
\end{equation}
and $\vec{y}_k=f[\vec{y}^{(1)},\cdots,\vec{y}^{(\alpha_k)}]$, $k=1,\cdots,r$, are linear functions of the derivatives of $\vec{y}$ with $\alpha_1<\cdots<\alpha_r$. The required highest order of differentiation, $\alpha_r$, is defined as the relative degree of the multi-input-multi-output system. The system is invertible if and only if the relative degree is finite.

The above inversion process also reveals that, to guarantee the transformation of observables exists, the number of measurement outputs must not be less than $m$. In practice, one can introduce redundant time-resolved measurements (i.e., $n>m$), which may reduce the risk of encountering singularity. This will be shown in the following numerical simulations.

\section{Numerical simulations}
\label{sec:simulation_result}
In this section, we carry out numerical simulations to show how the inversion-based algorithm can be applied for identifying {\it in situ} signals in quantum circuits.

Suppose that the system to be probed consists of two coupled transmon qubits with the following Hamiltonian \cite{Wu2018AnEA}:
\begin{equation}
H=\frac{\omega_1(t)}{2}\sigma_{1z}+\frac{\omega_2(t)}{2}\sigma_{2z} +g(t)(\sigma_{1+}\sigma_{2-}+\sigma_{1-}\sigma_{2+}),
\end{equation}
where $\sigma_z$ and $\sigma_\pm$ are the standard Pauli and lowering/raising operators.
The signals $\omega_1(t)$, $\omega_2(t)$ and $g(t)$ to be identified are the frequencies of the two qubits and the effective qubit-qubit coupling strength that are tunable by external electronic DC sources. In the following, we will apply the inversion algorithm to the identification of single-input and multi-input signals.

\subsection{Identification of single \textit{in situ} signals}

\begin{figure}[]
	\centering
	\includegraphics[width=1\columnwidth]{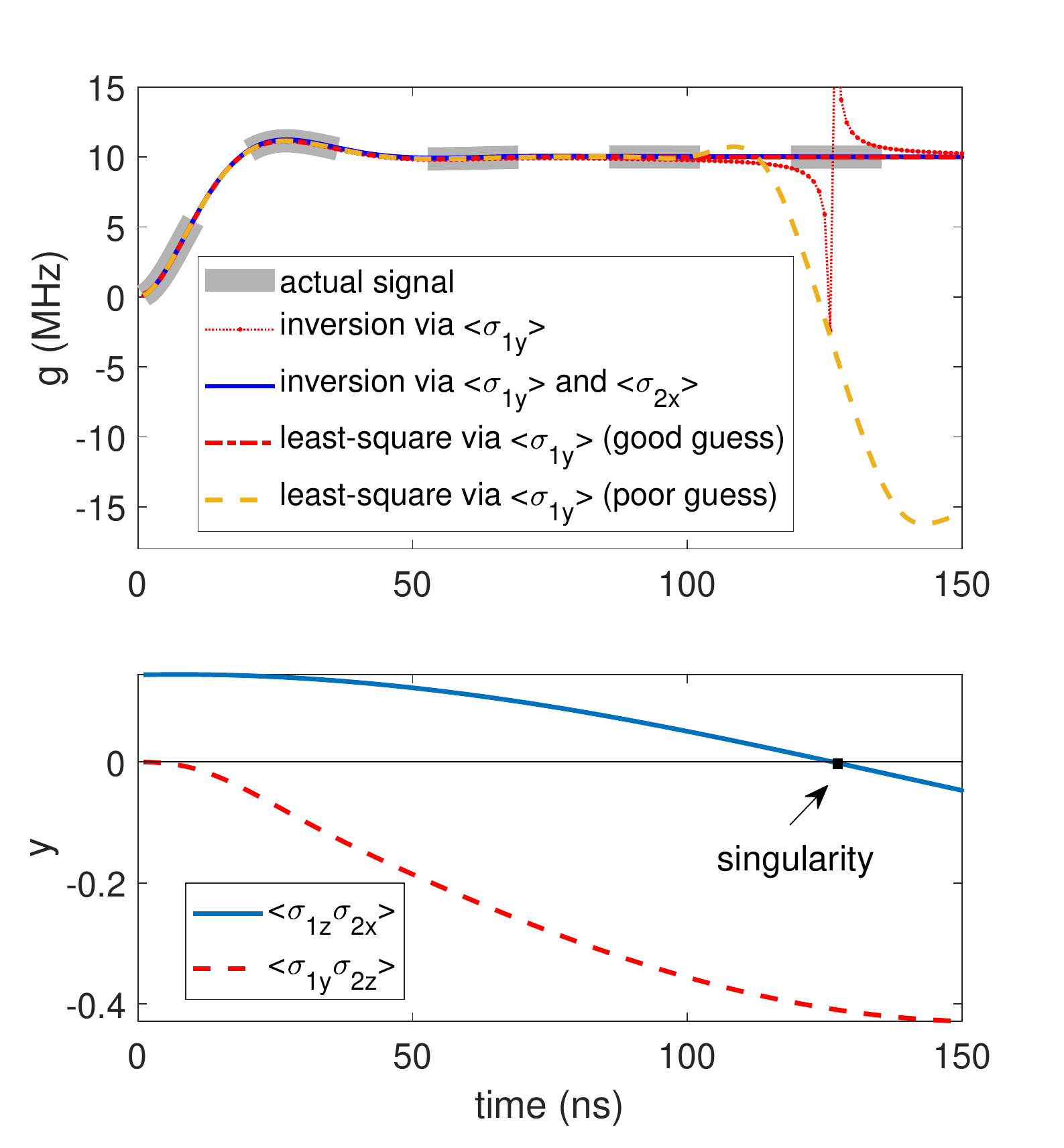}
	\caption{Identification of the {\it in situ} signal by using inversion method and least square method and the denominators in the inversion method. The identified signal by using inversion method by using single measurement and least-square method with a good and a bad initial guess on $g(t)$, respectively. The identification encounters singularity around $t=120$ns.  }
	\label{qubit1}
\end{figure}

In this case, we fix $\omega_1(t)=\omega_2(t)\equiv 1$MHz, and identify the time-varying $g(t)$ from time-resolved measurements. As shown in Fig.~\ref{qubit1}, the signal $g(t)$ to be identified is chosen to be a distorted step-function rising from $0$MHz to $10$MHz, which is often applied for quickly switching on the qubit-qubit coupling.

The initial state of the system is chosen to be
$$|\psi(0)\rangle=\left (\sqrt{\frac{2}{3}}|0 \rangle+\sqrt{\frac{1}{3}}|1 \rangle \right )\otimes \left (\frac{\sqrt{3}}{2}|0 \rangle +\frac{1}{2}|1 \rangle \right)$$
and the time-resolved measurement of $\sigma_{1y}$ is performed on the first qubit. As shown in Fig.~\ref{qubit1}, the identified signal $g(t)$ encounters singularity due to the vanishing of the denominator in
Eq.~(\ref{eq:InverseReadout}) at the critical time $t\approx 120$ns (shown in Fig.~\ref{qubit1}(b)), where the numerical simulation is instable and then comes back to the true solution. The inversion can be more stable (not shown here) by improving the numerical algorithm of solving the nonlinear differentia equation (e.g., using the three-point formula to approximate the differential of the measured ${y}(t)$ \cite{Burden2011}), but there is no guarantee that the inversion algorithm can always get over the singularity.

For comparison, we also apply the standard least-square method for the same problem \cite{Huang2012}, which searches the signal to be identified by minimizing the least-square error:
\begin{equation}
\mathcal{D}[g(t)]=\int_0^T \|y(t)-F(g(t))\|^2{\rm d}t,
\end{equation}
where $y(t)$ is the measured output and $F(u(t))$ is the output calculated through Eq.~(\ref{simple}) with input $g(t)$. Different from the inversion method, the least-square estimation needs to start from an initial guess on $g(t)$. We tested the method with a good guess $g^{(0)}(t)\equiv 10$MHz and a poor guess $g^{(0)}(t)\equiv-10$MHz, respectively. We find that the former converges to the true signal, but the latter diverges away just around the critical time.

Therefore, both the inversion and the least-square methods can well identify the {\it in situ} signal before the critical time, and may fail after it. The inversion algorithm is more efficient because it integrates the differential equation for only once without any initial guess on the signal, but the least-square method needs to solve the differential equation repeatedly until convergence.

The singularity cannot be removed by any particular identification algorithm. One must collect more information to uniquely determine the input signal, e.g., introducing redundant measurements. In the simulation, we introduce the second measurement $y'(t)=\langle\psi(t)|\sigma_{2x}|\psi(t)\rangle$  and the two denominators $\langle\psi(t)|\sigma_{1z}\sigma_{2x}]|\psi(t)\rangle$ and $\langle\psi(t)|\sigma_{1y}\sigma_{2z}|\psi(t)\rangle$ are not simultaneously zero. As is seen in Fig.~\ref{qubit1}, the inversion algorithm can precisely reproduce the {\it in situ} signal from redundant measurements.

\begin{figure}[]
	\centering
	\includegraphics[width=1\columnwidth]{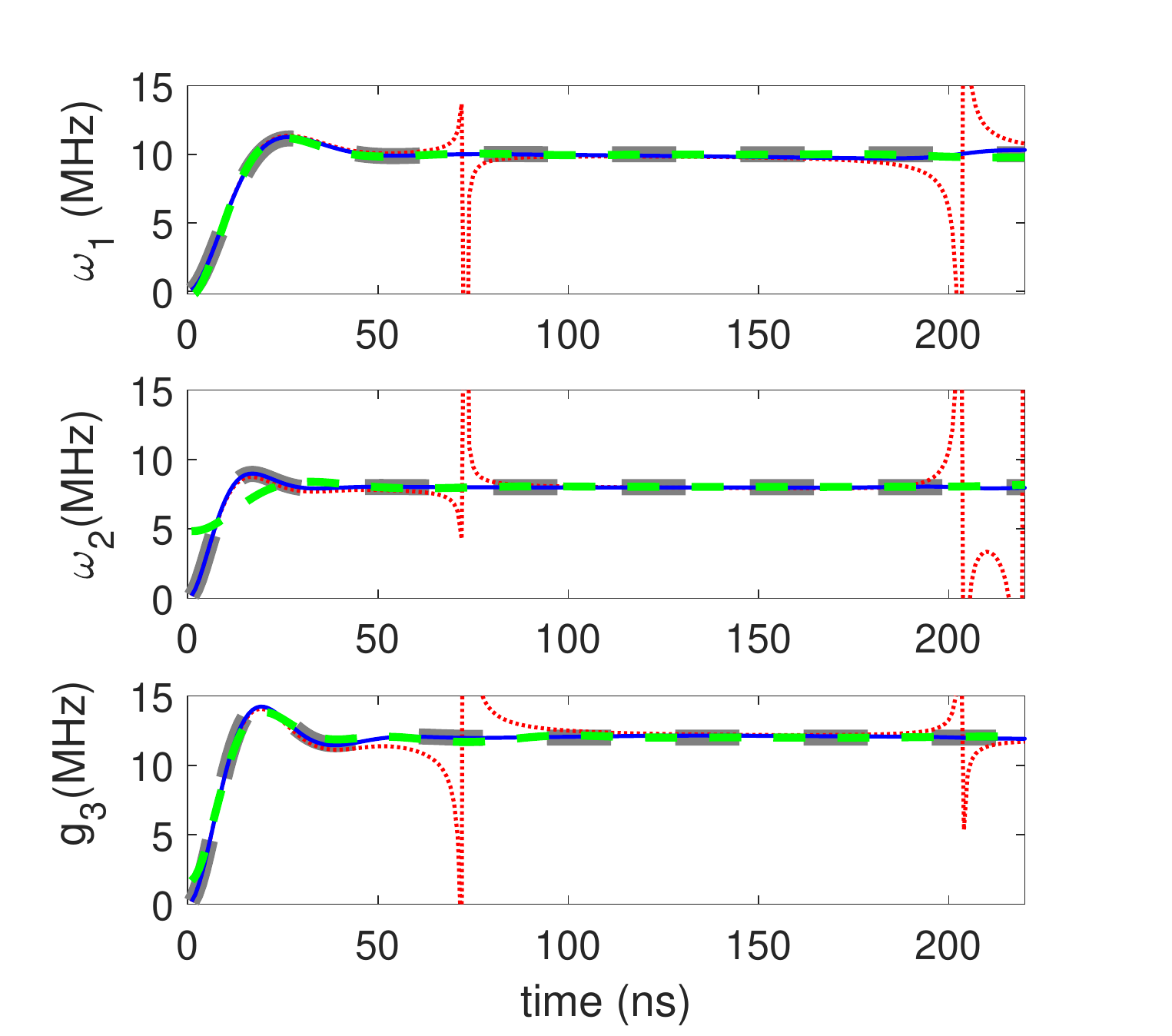}
	\caption{The identified multiple {\it in situ} signals $\omega_1(t)$ $\omega_2(t)$ and $g(t)$ (all chosen as distorted step functions) of the two-qubit system based on three (red dash line), five (blue solid line) time-resolved measurements and least-square method (green dash line). Singularity is present in the three-measurement case, but not in the five-measurement case.}
	\label{complex_system}
\end{figure}

\subsection{Simultaneous identification of multiple \textit{in situ} signals}
Assume that the {\it in situ} time-varying signals $\omega_1(t)$, $\omega_2(t)$ and $g(t)$ are all unknown. To identify them simultaneously, we need at least three time-resolved measurements, e.g.,
\begin{equation}\label{}
\vec{O}=(\sigma_{1x},\sigma_{1y},\sigma_{2x})^T,
\end{equation}
and the corresponding observable arrays $\vec{\mathcal{L}}_c^*{O}_1$,$\vec{\mathcal{L}}_c^*{O}_2$ and $\vec{\mathcal{L}}_c^*{O}_3$ can be examined to be linearly independent with each other. Hence, the system is invertible at least on a non-empty time interval with relative degree being $\alpha=1$.

We simulate the identification process with {\it in situ} signals $\omega_1(t)$, $\omega_2(t)$ and $g(t)$, which are all chosen as distorted step functions. As is shown in Fig.~\ref{complex_system}, the identified signals are all identical to the true ones until the first critical time $t\approx70$ ns. They come back to the true signals after a spiky deviation, and encounter again singularity at the second critical time $t\approx 200$ns. The singularities are removed by introducing two additional time-resolved measurements $O_4=\sigma_{2y}$ and $O_5=\sigma_{1z}$.

We also tested the least-square method for this case with initial guess $\omega_1^{(0)}(t)\equiv10$MHz, $\omega_2^{(0)}(t)\equiv10$MHz and $g^{(0)}(t)\equiv10$MHz, which are not far away from the actual signals. However, the resulting identified $\omega_2(t)$ fails to match the actual {\it in situ} signal, which corresponds to a trapping false solution. This implies that the least-square method in less reliable when multiple inputs are involved.

\subsection{The affection of noises}
The prevalently existing noises in realistic quantum systems can affect the quality of identification or even destroy it. Taking the one-qubit probe as an example, we consider two typical classes of noises present in the following system:
\begin{eqnarray}
\dot{\psi}(t) &=& -i\left[\textbf{n}_{\rm e}(t)\sigma_z+u(t)\sigma_z\right]\psi(t), \\
 y(t)& = &\langle \psi(t)|\sigma_x|\psi(t)\rangle+\n_{\rm m}(t).
\end{eqnarray}
The noise $\textbf{n}_{\rm e}(t)$ comes from unwanted coupling to unspecified signals (e.g., crosstalk via some other qubit's input), and $\textbf{n}_{\rm m}(t)$ comes from the imprecise measurement due to the randomness of quantum measurements or imperfect devices. According to Eq.~(\ref{eq:InverseReadout_simple}), the measurement noise, especially its high-frequency components, can have fatal effect on the quality of readout results because it can be greatly amplified by the differentiation of $y(t)$. The system is less affected by the high-frequency components of noise $\textbf{n}_{\rm e}(t)$ because they tend to be filtered.
\begin{figure}[ht]
	\centering
	\includegraphics[width=1\columnwidth]{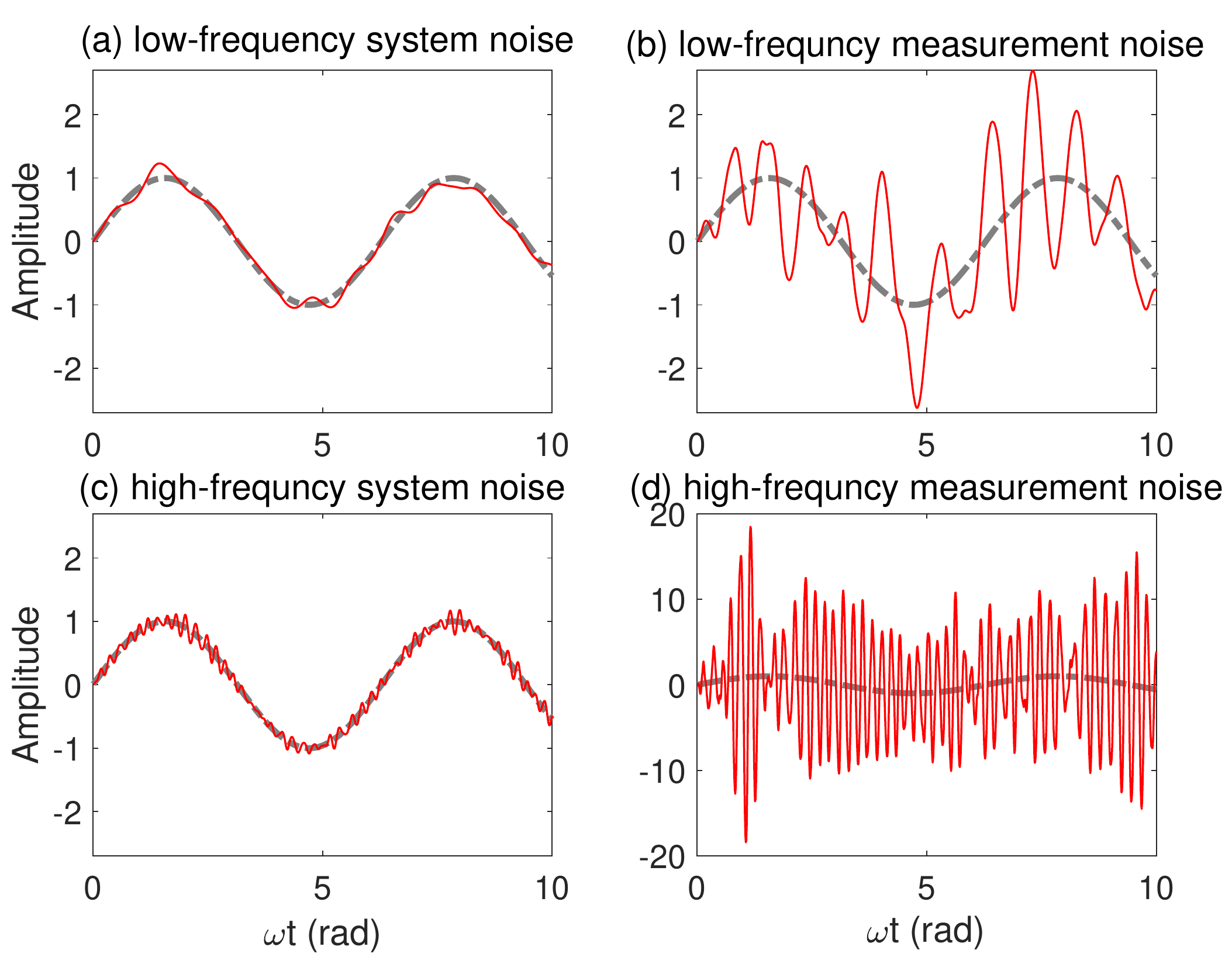}
	\caption{(a).The calculated \textit{in situ} signal when there is low frequency system noise ($\n_e$). (b).The calculated \textit{in situ} signal when there is high frequency system noise. (c).The calculated \textit{in situ} signal when there is low frequency measurement noise ($\n_m$). (d).The calculated \textit{in situ} signal when there is high frequency measurement noise.}
	\label{noise}
\end{figure}

In the simulation, we simulated low frequency (comparable with the frequency of the signals) and high frequency (25-30 times of the frequency of the signals) random noise in system ($\textbf{n}_e (t)$). The variance of the noise are taken to be $10^{-2}$ for $n_e(t)$ and $10^{-6}$ for $n_m(t)$, respectively. As shown in Fig.~\ref{noise} the measurement noise distorts the calculated \textit{in situ} signal dramatically even with a smaller variance, especially when the frequency is high. By contrast, the high frequency noises in the system has minor affection on the identification. Therefore, in practice, the measured output should be carefully filtered to reduce the noise affect while keeping the signal undistorted as much as possible.

\section{Conclusion}

To conclude, we propose the inverse-system based method to unambiguously identify time-varying \textit{in situ} signals from time-resolved measurements. Although the signals are usually locally identifiable (i.e., likely diverge at some critical time due to the singularity), the proposed method still greatly generalizes the existing Ramsey-experiment-based schemes to arbitrary multi-input-multi-output systems, as long as the algebraic invertibility condition is satisfied. The simulation results show that it can perfectly extract the \textit{in situ} signal in both single-input and multiple-input systems by integrating the nonlinear Schrodinger equation. Although, the algorithm is applicable only on a finite time interval due to potential singularity, one can properly introduce redundant measurements to prolong the applicable time interval. The affection and limitation brought by system's noises are also analyzed through numerical simulations.

The method we developed can be naturally generalized to any other quantum systems, no matter closed or open, as long as the probe system can be precisely modeled and the modeled system is invertible. In principle, one can freely choose the time-resolved measurements for extracting the {\it in situ} signals according to the invertibility condition. However, in practice one should pick those with lowest relative degree, so as to minimize the influence of measurement noise.
 
\bibliography{sample2}

\end{document}